\documentclass[review]{elsarticle}
\usepackage{graphicx}

\usepackage{amssymb}
\journal{J. Magn. Magn. Mater}

\begin{document}
\begin{frontmatter}
\title{First-principles prediction of spin-density-reflection symmetry driven magnetic transition of CsCl-type FeSe}
\author{Gul Rahman}
\author{In Gee Kim}
\ead{igkim@postech.ac.kr}
\address{Graduate Institute of Ferrous Technology, Pohang University of Science and Technology,\\
 Pohang 790-784, Republic of Korea}
\author{Arthur J. Freeman}
\address{Department of Physics and Astronomy, Northwestern University,\\
 Evanston, IL 60208, U. S. A.}
\begin{abstract}
Based on results of density functional theory (DFT) calculations with the local spin density approximation (LSDA) 
and the generalized gradient approximation (GGA), we propose a new magnetic material, CsCl-type FeSe. 
The calculations reveal the existence of ferromagnetic (FM) and antiferromagnetic (AFM) states over 
a wide range of lattice constants. At $3.12$\,{\AA} in the GGA, the equilibrium state is found to be AFM 
with a local Fe magnetic moment of $\pm 2.69\,\mu_\mathrm{B}$.
A metastable FM state with Fe and Se local magnetic moments of 
$2.00\,\mu_\mathrm{B}$ and $-0.032\,\mu_\mathrm{B}$, respectively,
lies $171.7$\,{meV} above the AFM state. Its equilibrium lattice constant is $\sim 2$\,{\%} smaller
than that of the AFM state, implying that when the system undergoes a phase transition from the AFM state
to the FM one, the transition is accompanied by volume contraction. 
Such an AFM-FM transition is attributed to spin-density $z$-reflection symmetry;
the symmetry driven AFM-FM transition is not altered by spin-orbit coupling.
The relative stability of different magnetic phases is discussed in terms of the local density of
states. We find that CsCl-type FeSe is mechanically stable, but the magnetic states are expected to be brittle.
\end{abstract}
\end{frontmatter}

\section{Introduction}
\label{sec:intro}

Iron-selenide shows a number of polymorphs including a low-temperature tetragonal ($\alpha$) phase, 
the so-called anti-PbO-type FeSe~\cite{ref-gul}. 
The $\alpha$-Fe$_\delta$Se$_{1-\delta}$ compound is currently
an interesting phase and has been studied experimentally for its spintronics related magnetic properties~\cite{Feng,Liu,Wu,Wu2}. 
The perfectly stoichiometric $\alpha$-FeSe has been shown to be non-ferromagnetic, 
but non-stoichiometric $\alpha$-FeSe demonstrates ferromagnetism 
that was attributed to defects or Fe clusters in $\alpha$-FeSe thin films~\cite{Wu2}. 
Zinc blende-type FeSe was shown to be an antiferromagnetic metal~\cite{gul-jkps}. 
Recently, superconductivity with a $T_c \sim 8$\,{K} was also discovered in $\alpha$-FeSe and 
was attributed to Se vacancies~\cite{Hu}. Subedi \textit{et al.}~\cite{Singh}
performed first-principles calculations and their results indicated that electron-phonon coupling 
can not explain the superconductivity of $\alpha$-FeSe which shows spin-density wave (SDW). 
Zhang \textit{et al.}~\cite{Zhang} showed that excess Fe in FeTe provides electron and
the excess Fe is strongly magnetic.

Motivated not only by the observed magnetism and superconductivity in $\alpha$-FeSe, 
but also by our recent extensive calculations of FeSe~\cite{ref-gul},
where we considered different crystal structures of FeSe and
{(i)} a phase transition from $\alpha$-FeSe to CsCl-type FeSe was found and
{(ii)} CsCl-type FeSe was the only structure that survived in a compressed unit cell and retained its magnetism. 
The tetragonal $\alpha$-FeSe structure with coordination number four can be thought of as the CsCl structure 
with an elongated $c$ axis and coordination number of eight. 
Since ordered CsCl-type systems have no nearest neighbor atoms of the same kind,
in order to understand the properties of such compounds containing magnetically active constituents, 
it is necessary to consider magnetic unit cells large enough to allow for antiferromagnetism.

Here, we investigate the magnetism of CsCl-type FeSe. 
We find that CsCl-type FeSe demonstrates both ferromagnetism and antiferromagnetism 
over a wide range of lattice constants in terms of density-functional theory (DFT)~\cite{DFT} by using the
total-energy all-electron full-potential linearized augmented plane-wave (FLAPW)~\cite{FLAPW} calculations implemented in the QMD-FLAPW software package~\cite{FLAPW-web}.

\section{Computational Method}
\label{sec:method}

Both the local spin density approximation (LSDA)~\cite{LSDA} and the generalized gradient approximation (GGA)~\cite{GGA} were considered for the exchange-correlation potentials. 
Integrations inside the Brillouin zone (BZ) were performed using the improved tetrahedron method~\cite{TETRA} 
over a $15\times 15\times 15$ mesh within the three dimensional (3D) BZ, 
corresponding to $120$\,{\textbf{k}} points inside the irreducible wedge of the 3D-BZ.
An energy cutoff at $4.1231$\,{$(2\pi/a)$}, where $a$ is the lattice constant, was employed for the linearized augmented plane-wave (LAPW) basis set, which corresponds to $\sim 210$\,{LAPWs} per \textbf{k}-point and spin. 
A $16.3707$\,{$(2\pi/a)$} star function cutoff was used for depicting the charge density and potential 
in the interstitial regions. 
Lattice harmonics with $l \leq 8$ were employed to expand the charge density, potential, and wave-functions 
inside each muffin-tin (MT) sphere of radius $2.2$\,{a.u.} for Fe and $1.9$\,{a.u.} for Se. 
Note that those computational parameters satisfy the convergence test~\cite{Seo}.

All core electrons were treated fully relativistically and valence states were calculated scalar relativistically, \textit{i.e.}, without spin-orbit coupling~\cite{Rela}.
The explicit orthogonalization (XO) scheme was employed to ensure the orthogonality between the core and valence states~\cite{XO}. For spin-orbit coupling on valence states, we employed the second variation method~\cite{SOC}
with the spin diagonal parts of the density subjected to a self-consistency loop. 
During the second variation procedure, integrations inside the 3D-BZ were done in the full-BZ, \textit{i.e.}, 
with 1688\,{\textbf{k}} points.
Self-consistency was assumed when the difference between input and output charge densities became less than $1.0\times10^{-4}$\,{electrons/a.u.$^{3}$}

\section{Results and Discussions}
\label{sec:results}
\subsection{Magnetic Phase Transition}
\label{ssec:m-transition}

\begin{figure}
\begin{center}
\includegraphics[]{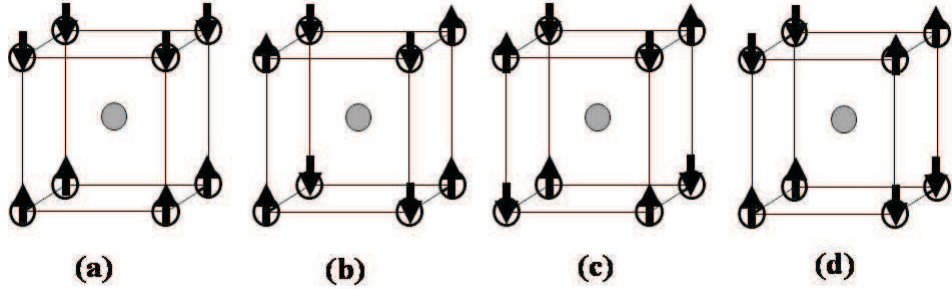}
\caption{The antiferromagnetic unit cells for (a) type-I, (b) type-II, (c) type-III, and (d) type-IV
presented on the basic CsCl unit cell for FeSe. 
Open and filled circles represent Fe and Se atoms, respectively. 
Arrows indicate different spin directions of the Fe atoms.} \label{structure}
\end{center}
\end{figure}

Because the LSDA and GGA results agree qualitatively with each other, 
we will keep our discussion based on the GGA results; 
the LSDA results will be given where they are required. 
The total energies were calculated for the nonmagnetic (NM) and ferromagnetic (FM) states 
using the conventional unit cell of two atoms.
Figure~\ref{structure} shows the antiferromagnetic (AFM) coupling of Fe in different planes and 
four types of AFM structures, AFM-I, AFM-II, AFM-III, and AFM-IV, were considered and 
tetragonally doubled unit cells were used for the AFM calculations. 
The main difference among these AFM structures is not only the AFM coupling 
between the Fe atoms in different planes, 
but also antiferromagnetically coupled coordination number($\nu$) in the unit cell 
\textit{e.g.}, the AFM-I, AFM-II, AFM-III, and AFM-IV states have two, four, six, and four antiferromagnetically coupled  atoms,
respectively. Furthermore, AFM-II and AFM-IV have the same coordination number, but the coupling between the Fe atoms is different when viewd in the $xy$ plane. Note that all these AFM structures are similar to the FeAs-based superconductor~\cite{FeAs}. 

\begin{figure}[h]
\begin{center}
\includegraphics[width=0.6\textwidth]{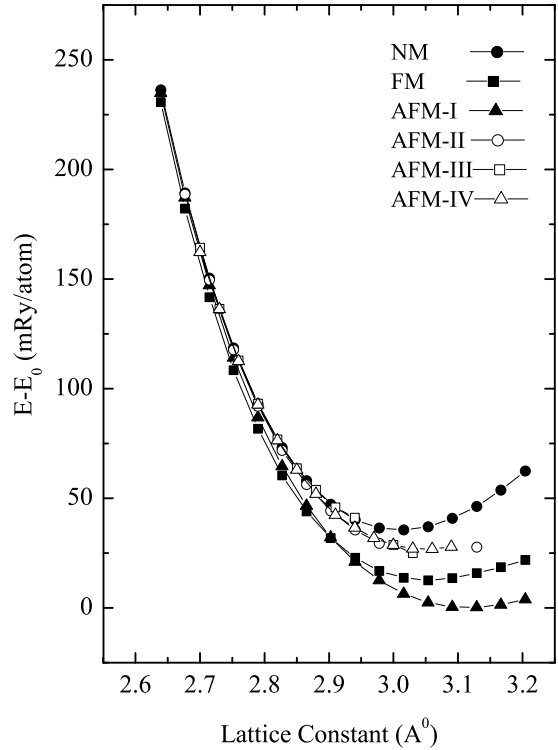}
\caption{The calculated total energy in units of {mRy/Fe atom} as a function of lattice constant (\AA) of 
FeSe in the NM, FM, AFM-I, AFM-II, AFM-III, and AFM-IV states relative to 
the equilibrium total energy of the AFM-I state.
Filled circles, squares, and triangles present the NM, FM, and AFM-I states, respectively. 
Open circles, squares, and triangles represent the AFM-II, AFM-III, and AFM-IV states of CsCl-type FeSe,
respectively.} \label{TEFIG}
\end{center}
\end{figure}

The calculated total energy energy curve shown in Fig.~\ref{TEFIG} demonstrates that 
the AFM-I state is the ground state of CsCl-type FeSe. 
The calculated total energies per Fe atom are shown relative to the equilibrium energy of the AFM-I state. 
The total energy difference between the FM and AFM states is sensitive to the unit cell volume change. 
Since the AFM-II, AFM-III, and AFM-IV states are higher in energy than the FM and AFM-I states, 
we denote for brevity hereafter the AFM-I state as the AFM state. 
The calculated equilibrium lattice constants for the NM, FM, and AFM states are summarized in Table~\ref{table-A}.
Calculations were also carried out for the optimization of the $c/a$ ratio of the AFM structure, 
and the optimized $c/a$ was found to be $\sim 1.985$.

\begin{table}
\caption{Calculated properties of CsCl-type FeSe; 
$a$ is the equilibrium lattice constant in {\AA} units, 
$m_\mathrm{Fe}$ is the local magnetic moment of Fe in $\mu_\mathrm{B}$ units, 
and $\Delta E$ gives the relative stability of the FM, AFM-II, and NM phases with respect to 
the AFM-I phase equilibrium in units of {meV/Fe atom}, \textit{i.e.}, 
$\Delta E < 0$ means that the AFM-I phase is more stable than  the FM, AFM-II, and NM phases. 
Results are given for LSDA and GGA.} \label{table-A}
\begin{tabular}{c|ccc|cccc}
\hline\hline \phantom{} & \multicolumn{3}{|c|}{LSDA} & \multicolumn{3}{c}{GGA} \\
\hline Phase& $a$ (\AA) &   $m_\mathrm{Fe}$ ($\mu_\mathrm{B}$) & $\Delta E$ (meV/Fe)
& $a$ (\AA) & $m_\mathrm{Fe}$ ($\mu_\mathrm{B}$) & $\Delta E$ (meV/Fe)\\
\hline
NM  & $2.94$ & $$ & $-229.97$ & $3.01$ & $$ & $-480.59$\\
FM & $2.96$ & $1.70$ & $\;\;-40.78$ & $3.05$ & $2.00$ & $-171.70$\\
AFM-I & $3.00$ & $2.12$ & $\;\;\;\;\;\;0.00$ & $3.12$ & $2.69$ & $\;\;\;\;\;\;0.00$\\
AFM-II & $2.96$ & $1.37$ & $-192.55$ & $3.05 $ & $ 2.09$ & $ -341.38$\\
\hline\hline
\end{tabular}
\end{table}

The energy difference per Fe atom between the AFM and FM states,
\begin{equation}
\Delta E=\frac{E(\mathrm{AFM})-E(\mathrm{FM})}{{n}}
\label{eq:energy-difference}
\end{equation}
where  $n$ is number of Fe atoms,
reflects the inter-atomic exchange coupling strength between Fe spins at the AFM equilibrium lattice, 
is calculated to be about $-205$\,{meV} ($-56$\,{meV}) with GGA (LSDA). 
The AFM state is more stable than the FM state for lattice constants larger than the critical value, 
$a_{c} = 2.90$\,{\AA}, where the AFM-FM transition occurs. 
It is to be mentioned that the LSDA gave $a_{c} = 2.92$\,{\AA}.  
The magnetic state of CsCl-type FeSe is very sensitive in the region $2.90$--$2.94$\,{\AA} 
and this is the region where FeSe makes a transition from the AFM state to the FM one under compression. The energy barrier ($\Delta E$) can be overcome if we decrease the lattice constant by a $\sim 7$\,{\%} ($\sim 2.6$\,{\%})($\frac{a_{eq}-a_{c}}{a_{eq}}\times100$) 
with GGA (LSDA) to result in the AFM CsCl-type FeSe in the FM state, 
as seen in Fig.~\ref{TEFIG}. 
To be more confident, we used the same tetragonal unit cell and found the FM stability against AFM 
in the region where the FM is more stable than the AFM.

It will be more interesting if we know how much external pressure is necessary 
to achieve this metastable CsCl-type FeSe. 
As we found, CsCl-type FeSe has different magnetic structures and one can calculate 
the transition pressures of these magnetic structures.
However, we only calculated the transition pressures for the NM, FM, and AFM states. 
The transition pressure was calculated by using the equation of state~\cite{Muri} which enables us to calculate the enthalpy $H = E + pV$, where $E$ is the internal energy, $p$ is the external pressure, and $V$ is the volume of the system. The transition pressure from $\alpha$-FeSe to CsCl-type FeSe is obtained from the usual condition of equal enthalpies, \textit{i.e.}, the pressure at which the enthalpies $H$ of $\alpha$-FeSe and CsCl-type FeSe are the same. We followed this approach and the transition pressure is determined by equating the enthalpies of the two phases of FeSe.
The GGA shows that pressure values of about $24.32$, $20.45$, and $19.90$\,{GPa} are
necessary for the transition of $\alpha$-FeSe to NM, FM, and AFM CsCl-type FeSe, respectively. 

\begin{figure}[t]
\begin{center}
\includegraphics[width=\textwidth]{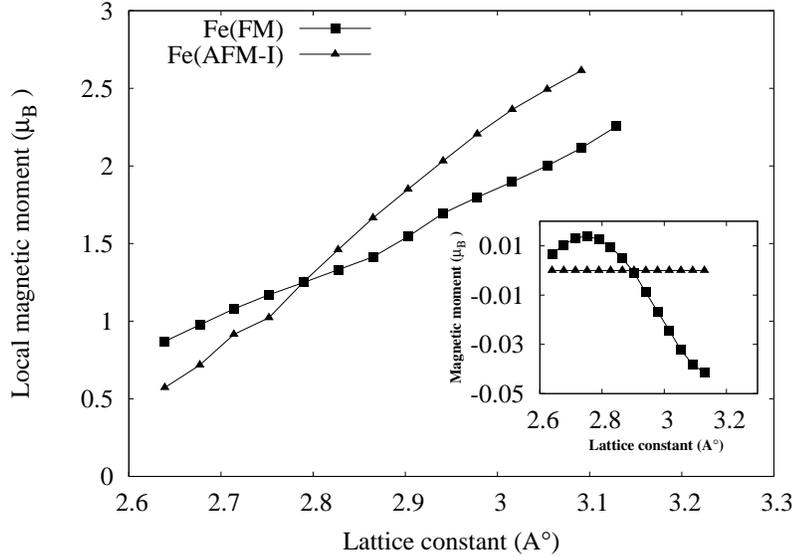}
\caption{Calculated local Fe magnetic moments (in $\mu_\mathrm{B}$ units) within each MT sphere. 
The inset shows the the local magnetic moments of Se for FM and AFM states. 
Squares and circles represent the local magnetic moment of Fe for the FM and AFM states, respectively.} 
\label{local-mm}
\end{center}
\end{figure}

The calculated local magnetic moments within each MT sphere of the Fe and Se atoms are shown 
in Fig.~\ref{local-mm} and the local Fe magnetic moments of all the magnetic phases are summarized 
in Table~\ref{table-A}. 
For both the FM and AFM states, the local magnetic moments of Fe increase as one increases the lattice constant. 
It is noticeable that there is a crossing point around $2.79$\,{\AA}. 
Below this lattice constant, where the FM state is more stable than the AFM one, 
the Fe local magnetic moments in the FM states are larger than those of the AFM state. 
Above $2.79$\,{\AA}, the Fe atom has a larger magnetic moment in the AFM state than in the FM state. 
The induced Se magnetic moment in the FM state was found to be 
$-0.032\,\mu_\mathrm{B}$ ($-0.012\,\mu_\mathrm{B}$) with GGA (LSDA). 
In Fig.~\ref{local-mm}, we can see that in the FM region the induced magnetic moments 
at the Se site are positive, whereas in the region where the AFM state is the most stable state, 
Se has negatively induced magnetic moments. 
A slight jump of the Fe moments in the FM state can be seen near the critical region of $2.90$--$2.94$\,{\AA}, 
where the transition takes place. 
These local magnetic moments show that the magnetic state (FM or AFM) is stabilized by 
the high spin state of the Fe atom.

\begin{figure}[h]
\begin{center}
\includegraphics[width=\textwidth]{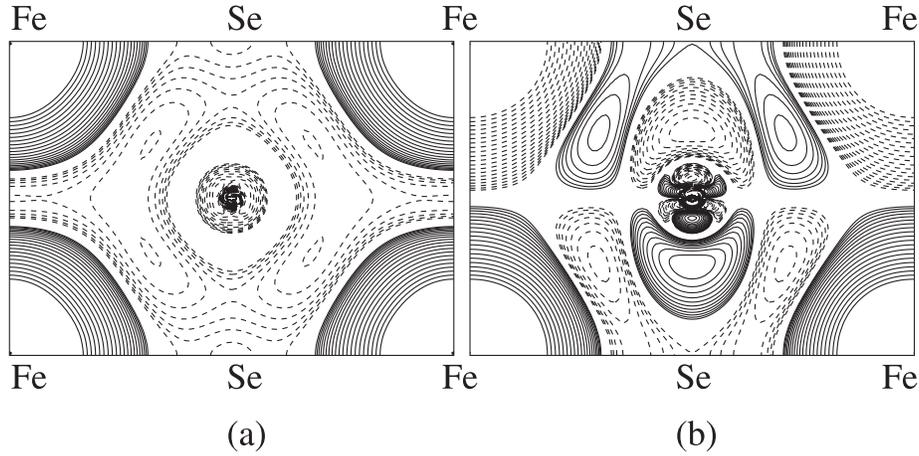}
\caption{Spin-density contours in the (110) plane of CsCl-type FeSe for the  (a) FM and (b) the AFM states. 
The solid and broken lines represent spin-up and spin-down densities, respectively. 
The lowest contour starts from $2\times 10^{-4}$\,{electrons/a.u.$^3$} 
and the subsequent lines differ by a factor of $\sqrt{2}$. 
The Fe and Se atoms are located at the corner and center of each panel, respectively.} \label{spindensity}
\end{center}
\end{figure}

It is interesting to find that the calculated local magnetic moment of the Se atom in the AFM state is zero. 
To understand this feature, the spin-density contour plots of CsCl-type FeSe are shown 
in Fig.~\ref{spindensity} for both the FM and AFM states. 
The FM state shows, for the spin-density $z$-reflection operation, even symmetry, 
while the AFM state shows odd symmetry. 
This implies that the FM-AFM transition should be a first-order phase transition, 
because a first-order phase transition is understood as the choice of 
the representation of a symmetry operation~\cite{Landau}. 
Kim \textit{et al.}~\cite{igkim} revealed the spin-density-inversion symmetry driven 
first-order FM-AFM phase transition of GaCMn$_3$. 
Considering the fact that the inversion operation can be understood as a rotation followed by a reflection, 
it is possible to generalize this idea to the spin-density-reflection symmetry driven 
first-order FM-AFM phase transition.

In order to check the effects of SOC, we also carried out calculations in the FM and AFM states with SOC. 
However, there is no significant alteration of the symmetry driven transition by SOC.

\subsection{Electronic Structures}
\label{ssec:e-structure}

\begin{figure}[t]
\begin{center}
\includegraphics[width=0.30\textwidth]{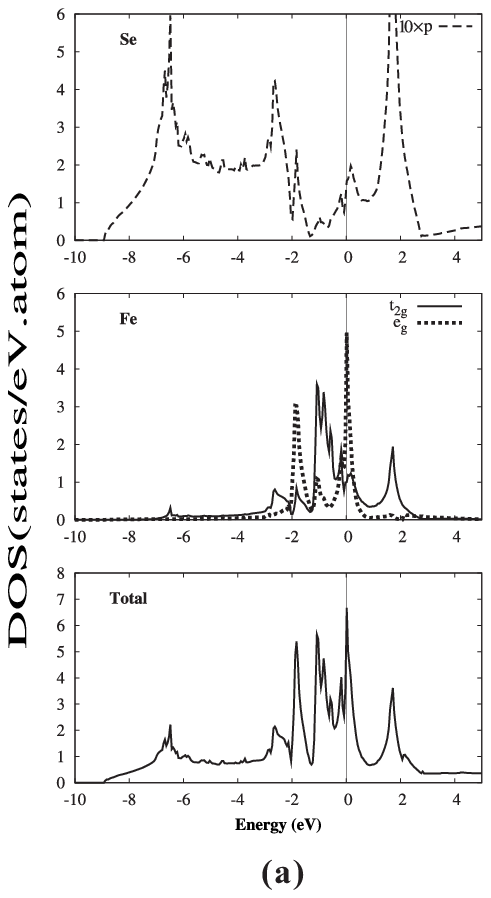}
\includegraphics[width=0.30\textwidth]{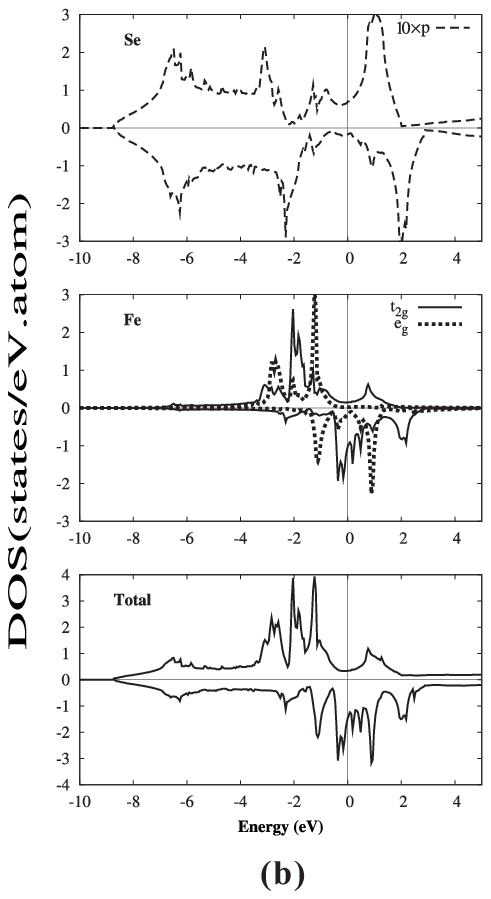}
\includegraphics[width=0.30\textwidth]{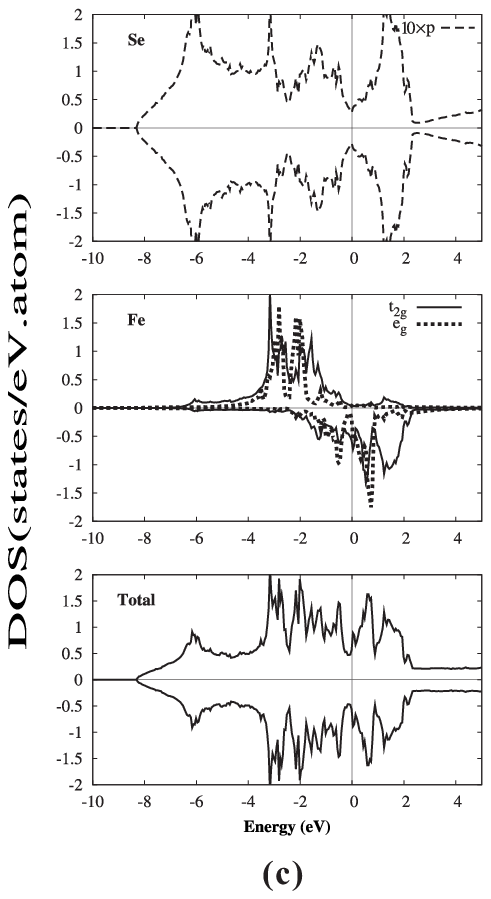}
\caption{Total and atom-projected density of states of the 
(a) NM, (b) FM, and (c) AFM states of CsCl-type FeSe at their equilibrium lattice constants. 
Solid, dotted, and dashed lines show Fe $t_{2g}$, $e_{g}$ and Se $p$ states, respectively, 
whereas solid lines in the bottom panels show the total DOS. 
The Se $p$ states are multiplied by a factor of 10 and the total density of states of AFM FeSe is
given per formula unit for comparison purposes. 
The Fermi ($E_\mathrm{F}$) levels are set to zero.} \label{dosfig}
\end{center}
\end{figure}
The relative stability among the magnetic orderings can be understood from an analysis of 
the calculated density of states (DOS). 
Figure~\ref{dosfig}(a) shows the DOS for the NM state of CsCl-type FeSe; 
the Fe $d$ states are decomposed into $t_{2g}$ and $e_g$ states. 
Note that the high peak at the Fermi level ($E_\mathrm{F}$) originates from Fe-$e_g$ states hybridized 
with the Se-$p$ states. 
The large DOS at $E_\mathrm{F}$, which comes mainly from Fe-$e_g$ states, 
indicates the ferromagnetic instability in terms of the mean-field Stoner theory~\cite{meanfield}. 
One may observe that the $t_{2g}$ states lie between the $e_{g}$ states. 
Such  behavior is due to the structure of CsCl-type FeSe which is typical for body-centered cubic metals 
where a pseudogap separates the bonding and anitbonding states~\cite{Book-ref}. 
Such a splitting can also be seen in the FM DOS. 
Once the FM ordering happens, as in Fig.~\ref{dosfig}(b), one can see the exchange splitting in the Fe-$d$ states,
along with the fact that the Fe-$e_{g\uparrow}$ states are occupied completely, 
while the Fe-$t_{2g\uparrow}$ states are partially occupied. 
Strong hybridizations between the Fe-$d$ states and the Se-$p$ states are also observed in both spins.

An important difference between the FM and AFM structure is that in FM the effective field split the bands and 
hence the peaks in the total DOS, while in the AFM case as seen in Fig.~\ref{dosfig}(c), 
the spin-up and spin-down peaks do not exhibit any such splitting and differ only in their relative intensity. 
An additional feature of the AFM phase is the absence of Se local magnetic moments consistent with 
the spin density maps. 
The populations of the partial states of Se atoms are very small in both magnetic structures, 
therefore the Fe $d$ states which are dominant at $E_\mathrm{F}$ and 
behave differently in the FM and AFM phases are primarily responsible for their relative stability. 
Furthermore, in the AFM structure, $E_\mathrm{F}$ does not fall in the valley, 
the majority Fe-$t_{2g}$ and Fe-$e_g$ states are completely occupied, 
and the DOS at $E_\mathrm{F}$  is very small and leads to a stable AFM structure, 
while the FM splitting leads to an increased DOS at $E_\mathrm{F}$  mainly contributed by Fe Fe-$t_{2g}$. 
In addition, the Fe-$d$ spin-up states are mostly occupied and 
provide the relatively higher local Fe magnetic moment to be $2.69$\,{$\mu_\mathrm{B}$}, 
which is larger than that of the FM case of $2.00$\,{$\mu_\mathrm{B}$}. 
state. 
It is to be noted that $\Delta E$, $m_\mathrm{Fe}$, and density of states at $E_\mathrm{F}$ of CsCl-type FeSe 
are larger than those in $\alpha$-FeSe~\cite{Singh}.

\subsection{Mechanical Stability}
\label{ssec:mech}

\begin{figure}
\begin{center}
\includegraphics[width=\textwidth]{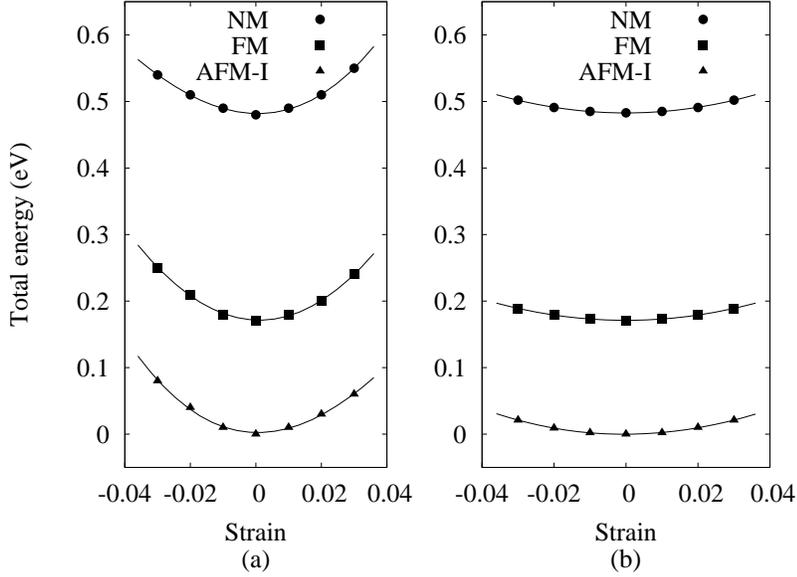}
\caption{Total energy per Fe atom (in units of eV) versus 
(a) tetragonal and (b) orthorhombic strain of FeSe in their NM, FM, and AFM phases. 
The total energy is given with respect to unstrained AFM FeSe for tetragonal and orthorhombic distortion. 
Filled solids, squares, and triangles represent the NM, FM, and AFM-I states, respectively.}
\label{strain-fig}
\end{center}
\end{figure}

Finally, an important question to address is  whether this new material is mechanically stable or not. 
To answer this question we calculated the tetragonal and trigonal shear constants, 
$C^{\prime}$ and $C_{44}$, respectively of the NM, FM, and AFM states, 
by computing the change in energy of the FeSe phase under small volume-conserving strains. 
The results of tetragonal and orthorhombic strains are shown in Fig.~\ref{strain-fig} and 
the calculated values are summarized in Table~\ref{table-elastic}. 
Considering the fact that CsCl-type FeSe is metastable~\cite{ref-gul}, it is mechanically stable 
in the NM, FM, and AFM phases against tetrahedral and orthorhombic deformations. 
The mechanical stability is confirmed by the positive value of the cubic elastic constant 
$C^{\prime}=(C_{11}-C_{12})/2$ shown in Table~\ref{table-elastic}. 
However, one should care on the brittle nature of the magnetic states as represented by 
the shear modulus~\cite{YWu} on
the slip plane $\mu = \left( 3C_{44} + C_{11} - C_{12}\right)/5$, 
to the bulk modulus $B$ in Table~\ref{table-elastic}, 
where all the magnetic states are predicted to be brittle $\mu / B > 0.5$.
Note that the GGA predicts higher brittleness of the FM state than the LSDA;
this fact reflects the difference in bonding nature of the approximations.
It is interesting to find that the NM state is also mechanically stable and rather ductile, 
\textit{i.e.}, $\mu / B < 0.4$, in contrast to bcc Fe~\cite{gul-bccFe}. It is to be noted that the detailed mechanical behavior of CsCl-type FeSe at high pressure (reduced lattice parameters) is beyond the scope of the current study. At reduced lattice parameters, the nature of the brittleness of CsCl-type FeSe may change.

\begin{table}
\caption{The calculated bulk modulus ($B$), shear moduli ($C^{\prime}$ and $C_{44}$), 
and shear moduli on the slip plane ($\mu$) in GPa units of CsCl-type FeSe in 
the NM, FM, and AFM ($=$AFM-I) states  for both LSDA and GGA.} \label{table-elastic}
\begin{tabular}{c|ccccc|ccccc}
\hline\hline \phantom{} & \multicolumn{5}{|c|}{LSDA} & \multicolumn{5}{c}{GGA} \\
\hline Phase& $B$ &   $C^{\prime}$& $C_{44}$ & $\mu$ & $\mu/B$ 
& $B$ & $C^{\prime}$ & $C_{44}$ & $\mu$ & $\mu/B$\\
\hline
NM  & $174.6$ &  $74.7$ & $73.8$ & $74.2$ & $0.42$ & $142.6$ &  $66.6$ & $61.4$ & $63.5$ & $0.45$\\
FM &   $146.6$ &  $99.4$ & $74.0$ & $84.2$ & $0.57$ & $\;\;93.0$ &  $80.1$ & $55.2$ & $65.2$ & $0.70$\\
AFM & $117.0$ & $90.3$ & $82.5$ & $85.6$ & $0.73$ & $\;\;82.7$ & $60.3$ & $61.7$ & $61.1$ & $0.74$\\
\hline\hline
\end{tabular}
\end{table} 

The lattice constant over which the FM state becomes stable is close to that of FM bcc Fe, $2.87$\,{\AA}, 
which has been grown on GaAs~\cite{GaAs, GaAs1} and ZnSe~\cite{refZnSe,refZnSe1} substrates 
due to the good lattice match between them. 
Therefore, it can also raise the possibility of growing CsCl-type FeSe on GaAs or ZnSe, 
because the calculated lattice parameter of FeSe is very close to that of GaAs 
($a_{\mathrm{AFM}} \cong a_{\mathrm{GaAs}}/2$). 
The formation energy of CsCl-type FeSe at ambient pressure is $\sim 1$\,{eV/f.u.}~\cite{ref-gul}, 
which indicates that CsCl-type FeSe is in a metastable state and 
there might be a possibility to achieve this structure either at high pressure or grown as a thin film. 
Recently, FeS, which is isoelectronic to FeSe, indeed showed CsCl structure at high pressure~\cite{ref-FeS}. 
Several other metastable materials have been grown successfully~\cite{CrAs,CrAs1,CrSb,bccNi,CrTe} 
using molecular beam epitaxy (MBE), which is a non-equilibrium process.

\section{Summary}
\label{sec:summary}
In summary, we predict, based on density functional theory by using the total-energy all-electron FLAPW method, 
a new phase of FeSe, the FM and AFM state of CsCl-type FeSe . 
The calculated total energy curves indicate that the AFM state is most stable with the largest lattice constant. 
In addition, the energy barrier for the AFM to FM transition can be overcome by volume contraction. 
We found a spin-density-reflection symmetry driven first-order phase transition in CsCl-type FeSe, 
as seen previously in the spin-density-inversion symmetry driven phase transition in GaCMn$_3$~\cite{igkim}. 
The calculated LDOS and spin-density contour plots revealed the origin of the magnetic phase transitions 
in CsCl-type FeSe. 
We also predict that CsCl-type FeSe is mechanically stable for all the magnetic states considered in this work.

\section*{Acknowledgements}

The authors appreciate H. K. D. H. Bhadeshia for his careful reading of the manuscript. 
I. G. Kim thanks Jong-Hoon Chung for helpful discussions on experimental realization. 
This work was supported in part by the Steel Innovation Program of POSCO,
by the Basic Science Research Program (Grant No. 2009-0088216) through 
the National Research Foundation funded by 
Ministry of Education, Science and Technology of the Republic of Korea,
and by the U.S. Department of Energy (Grant No. DE-FGO2-88ER 45372).

\end{document}